\documentclass[11pt,preprint]{emulateapj}
\usepackage{graphicx}

\begin{document}

\title{ALMA Observation of the 658~GHz Vibrationally Excited H$_{2}$O Maser in Orion~KL Source~I}

\author{Tomoya HIROTA\altaffilmark{1,2}, 
Mi Kyoung KIM\altaffilmark{3}, 
Mareki HONMA\altaffilmark{1,2}} 
\email{tomoya.hirota@nao.ac.jp}
\altaffiltext{1}{National Astronomical Observatory of Japan, Osawa 2-21-1, Mitaka, Tokyo 181-8588, Japan}
\altaffiltext{2}{Department of Astronomical Sciences, SOKENDAI (The Graduate University for Advanced Studies), Osawa 2-21-1, Mitaka, Tokyo 181-8588, Japan}
\altaffiltext{3}{Korea Astronomy and Space Science Institute, Hwaam-dong 61-1, Yuseong-gu, Daejeon, 305-348, Republic of Korea}

\begin{abstract}
We present an observational study of the vibrationally excited H$_{2}$O line at 658~GHz ($\nu_{2}$=1, 1$_{1, 0}$-1$_{0, 1}$) toward Orion~KL using the Atacama Large Millimeter/Submillimeter Array (ALMA). 
This line is clearly detected at the position of the massive protostar candidate, the Source~I. 
The spatial structure is compact with a size of about 100~AU and is elongated along the northeast-southwest low-velocity  (18~km~s$^{-1}$) bipolar outflow traced by 22~GHz H$_{2}$O masers, SiO masers, and thermal SiO lines. 
A velocity gradient can be seen perpendicular to the bipolar outflow. 
Overall spatial and velocity structure seems analogous to that of the 321~GHz H$_{2}$O maser line previously detected with ALMA and vibrationally excited SiO maser emission. 
The brightness temperature of the 658~GHz H$_{2}$O line is estimated to be higher than 2$\times$10$^{4}$~K, implying that it is emitted via maser action. 
Our results suggest that the 658~GHz H$_{2}$O maser line is emitted from the base of the outflow from a rotating and expanding accretion disk as observed for the SiO masers and the 321~GHz H$_{2}$O maser. 
We also search for two other H$_{2}$O lines at 646~GHz (9$_{7, 3}$-8$_{8, 0}$ and 9$_{7, 2}$-8$_{8, 1}$), but they are not detected in Orion~KL. 
\end{abstract}


\section{Introduction}

The Orion Kleinmann-Low \citep[KL;][]{kleinmann1967} region is known as the nearest site of active massive-star formation at a distance of 420~pc \citep{menten2007, kim2008}. 
There are number of compact sources possibly associated with newly born young stellar objects (YSOs). 
The most dominant energy source is thought to be a radio source identified as Source~I \citep{churchwell1987,menten1995}. 
Because of its extremely high opacity at infrared wavelengths \citep{greenhill2004a, sitarski2013}, observational studies on Source~I have been made mainly from centimeter to submillimeter wavelengths. 

Observations with the Very Large Array (VLA) at 43~GHz resolve the continuum emission from Source~I into a compact structure elongated in northwest-southeast direction with a major axis of $\sim$100~AU  \citep{reid2007, goddi2011}. 
This emission can be explained by i) H$^{-}$ free-free radiation from optically thick neutral gas at $<$4500~K heated via mass accretion or ii) proton-electron free-free emission from compact H{\sc{ii}} region or from an ionized accretion disk with a higher temperature of $\sim$8000~K \citep{beuther2006, reid2007, goddi2011, plambeck2013, hirota2015}. 
Recent studies favor the H$^{-}$ free-free radiation scenario \citep{reid2007, plambeck2013, hirota2015}, while a luminosity of Source~I required for the dynamical interaction scenario (see below) prefers the proton-electron free-free emission \citep{goddi2011}. 
Future higher resolution observations at higher frequencies are required to decide between the two possibilities. 

Proper motion measurements of Source~I and the BN object, another high-mass YSO, with VLA have revealed that they are moving away from each other at a relative velocity of $\sim$40~km~s$^{-1}$ \citep{gomez2008, goddi2011}. 
It is proposed that Source~I and BN were ejected from a multiple system due to a dynamical interaction that occurred about 500~yrs ago \citep{goddi2011, bally2011}, and Source~I formed a tight binary system with a total mass of 20$M_{\odot}$. 
At the same time, the dynamical decay released gravitational energy and ejected an explosive wide-angle outflow at an expansion velocity of $>$100~km~s$^{-1}$ traced by the shocked H$_{2}$ and [Fe {\sc{ii}}] lines \citep{allen1993, kaifu2000, bally2011, bally2015}, which largely provides the $10^{5}~L_{\odot}$ luminosity of the KL nebula. 
It powers the strong (sub)millimeter line emission from the famous ``hot core'' \citep{beuther2008, zapata2009b, zapata2011}. 
On the other hand, there is an alternative explanation for the origin of the proper motions of this system \citep{tan2004, chatterjee2012}, in which the BN object was ejected 4500~yr ago by a dynamic interaction with the $\theta^{1}$C system in the Trapezium cluster rather than that with Source~I. 
Therefore, such a dynamical interaction scenario is still a matter of debate. 

Source~I is also known as a driving source of another bipolar outflow along the northeast-southwest direction. 
The inner most part of the outflow at 10-100~AU scales is traced by vibrationally excited ($v$=1, 2) SiO masers \citep{menten1995, wright1995, baudry1998, greenhill1998, doeleman1999, doeleman2004, greenhill2004b, kim2008, goddi2009, matthews2010, niederhofer2012a, greenhill2013}. 
Source~I has been recognized as one of three rare YSOs displaying vibrationally excited SiO masers \citep{morita1992, zapata2009a}, suggesting a very large luminosity of $>$10$^{4}L_{\odot}$, as expected for late-type stars associated with SiO masers \citep{menten1995}. 
VLBI observations have revealed that the vibrationally excited SiO masers at 43~GHz ($J$=1-0) trace the limb of an expanding, rotating disk see edge-on that appears to be the base of a magnetically driven disk wind \citep{greenhill2004b, matthews2010, greenhill2013}. 
They show a velocity gradient perpendicular to the outflow suggesting a rotating structure with an enclosed mass of $\sim$7-8$M_{\odot}$ \citep{kim2008, matthews2010}. 
This value is roughly consistent with that inferred from infrared spectroscopy of Source~I, which suggests a 10$M_{\odot}$ protostar with a circumstellar disk \citep{testi2010}, while it is significantly lower than that proposed by the dynamical interaction scenario, a 20$M_{\odot}$ binary \citep{goddi2011, bally2011}. 
The smaller mass estimates may provide a lower limit, if there is non-gravitational support such as radiation pressure and magnetic forces \citep{matthews2010}. 

Outside the vibrationally excited SiO maser distribution, the outflow can be traced by ground vibrational state ($v$=0) SiO masers and 22~GHz H$_{2}$O masers on scales of 100-1000~AU \citep{genzel1981, gaume1998, hirota2007, greenhill2013, neufeld2013}. 
Proper motion measurements of these maser features provide a 3-dimensional velocity field for the outflow with an expansion speed of $\sim$18~km~s$^{-1}$ \citep{genzel1981, greenhill2013}. 
This northeast-southwest outflow is sometimes called low-velocity (18~km~s$^{-1}$) outflow. 
The outflow lobes continue to extend to larger than 1000~AU scales, traced by thermal SiO emission \citep{wright1995, plambeck2009, zapata2012, niederhofer2012a} interacting with the ambient gas. 

Recent observations with the Atacama Large Millimeter/Submillimeter Array (ALMA) have detected submillimeter H$_{2}$O lines at 321~GHz ($10_{2,9}$-$9_{3,6}$) and 336~GHz ($\nu_{2}$=1, 5$_{2,3}$-$6_{1,6}$) toward Source~I \citep{hirota2014a}. 
Maps of both these lines show velocities that are consistent with those of the SiO masers, i.e., a velocity gradient perpendicular to the bipolar outflow as seen in the SiO maser distributions. 
While the 321~GHz line emission is elongated along the bipolar outflow, the 336~GHz vibrationally excited line emission is unresolved with the ALMA beam size of 0.4\arcsec. 
The velocity centroid map of the 336~GHz line is consistent with an origin in an edge-on rotating disk described above, i.e. a diameter of 0.2\arcsec \ (84~AU) and an enclosed mass of $>$7$M_{\odot}$. 
The spectral profile of the 336~GHz line can be explained with an excitation temperature of $>$3000~K, which is consistent with its lower state energy, $E_{l}$, of 2939~K. 
Thus, the 336~GHz line is consistent with thermal excitation, whereas the 321~GHz line could be a maser \citep{alcolea1993}. 
ALMA science verification (SV) data has also been used to detect the higher energy transition at 232~GHz ($\nu_{2}$=1, 5$_{5,0}$-6$_{4,3}$) at a lower state energy level of $E_{l}$=3451~K \citep{hirota2012}. 
Although the spatial and velocity structure of the 232~GHz line was not resolved, due to its insufficient spatial resolution, detection of the high energy H$_{2}$O transition would imply evidence for a hot molecular gas component very close to the protostar candidate Source~I \citep{reid2007, testi2010}. 

The results discussed above demonstrate that the submillimeter H$_{2}$O lines at high excitation energy can play complementary roles to various SiO thermal and maser lines and the 22~GHz H$_{2}$O maser, providing physical and dynamical properties of the hot molecular gas associated with newly born YSOs on a scale of $\sim$100~AU. 
However, the physical properties and dynamical structure of the H$_{2}$O emission region have not been fully understood yet, owing to the limited amount high resolution data \citep{hunter2007, patel2007, hirota2012, niederhofer2012b, hirota2014a, richards2014}. 

In this paper, we report ALMA observations of the vibrationally excited H$_{2}$O line at 658~GHz toward Orion~KL Source~I. 
The 658~GHz H$_{2}$O line is the lowest rotational transition at the first vibrationally excited state of the ortho H$_{2}$O ($\nu_{2}$=1, 1$_{1, 0}$-1$_{0, 1}$), with a lower state energy level of $E_{l}$=2329~K \citep{chen2000}. 
The 658~GHz H$_{2}$O maser lines has been detected toward late-type stars by the Caltech Submillimeter Observatory (CSO) 10.4~m telescope \citep{menten1995b}, the Submillimeter Array (SMA) \citep{hunter2007}, and the recently released ALMA SV data \citep{richards2014}.  However, it is not seen toward the star-forming regions W49N and W51 IRS2 \citep{menten1995b}, in spite of their extremely strong 22~GHz H$_{2}$O maser emissions. 
The 658~GHz H$_{2}$O line was identified in a single-dish line survey toward the Orion~KL with the CSO 10.4~m telescope \citep{schilke2001}. 
However, its position and source size could not be determined, because of the large beam size of the single-dish telescope $\sim$10\arcsec. 
We present high resolution imaging of the 658~GHz H$_{2}$O lines for the first time in star-forming region with a spatial resolution of $\sim$0.26\arcsec \ or 100~AU. 
The compact structure and high flux density suggest that the 658~GHz line is showing maser action. 
Given that SiO and H$_{2}$O are the only molecules from which emission has been detected toward Source~I, further high-resolution, multi-transition studies of vibrationally excited H$_{2}$O lines \citep[e.g.][]{humphreys2007} will greatly contribute to our understanding of the structure and dynamics of disk/outflow systems associated with this highly interesting object. 

\section{Observations}

\begin{figure}
\begin{center}
\includegraphics[width=8cm]{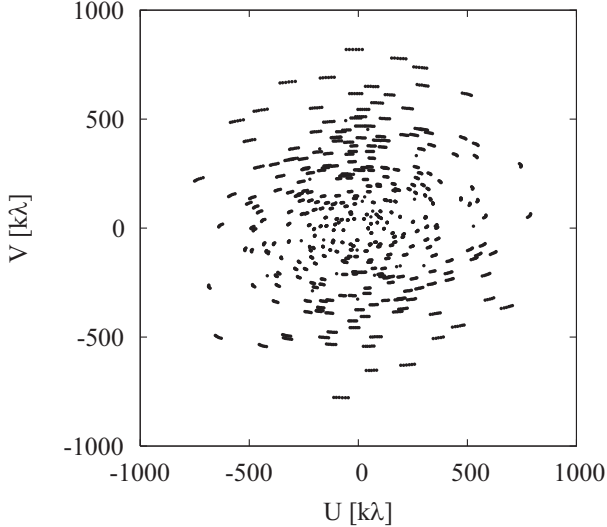}
\caption{UV coverage of the ALMA observation at band 9 on Aug. 25, 2012. }
\label{fig-uv}
\end{center}
\end{figure}

Observations with ALMA at band 9 were carried out during Cycle 0 at five epochs (ADS/JAO.ALMA\#2011.00199.S). 
Among them, we selected the Aug. 25, 2012 observation for analysis in the present paper, as it contained the most antennas, longeset baselines, and most on-source time. 
Total on-source time was 350~seconds, consisting of 6 scans of 30 or 72~seconds every 10~minutes, and the maximum baseline length was 385~m (845~k$\lambda$). 
The UV coverage for this session is shown in Figure \ref{fig-uv}. 
Because of the lack of visibility data at 658~GHz and/or large flux uncertainties from epoch to epoch as discussed later, we did not combine multiple epoch data. 
Typical system noise temperatures were about 1400~K at 658~GHz. 

\begin{figure*}
\begin{center}
\includegraphics[width=16cm]{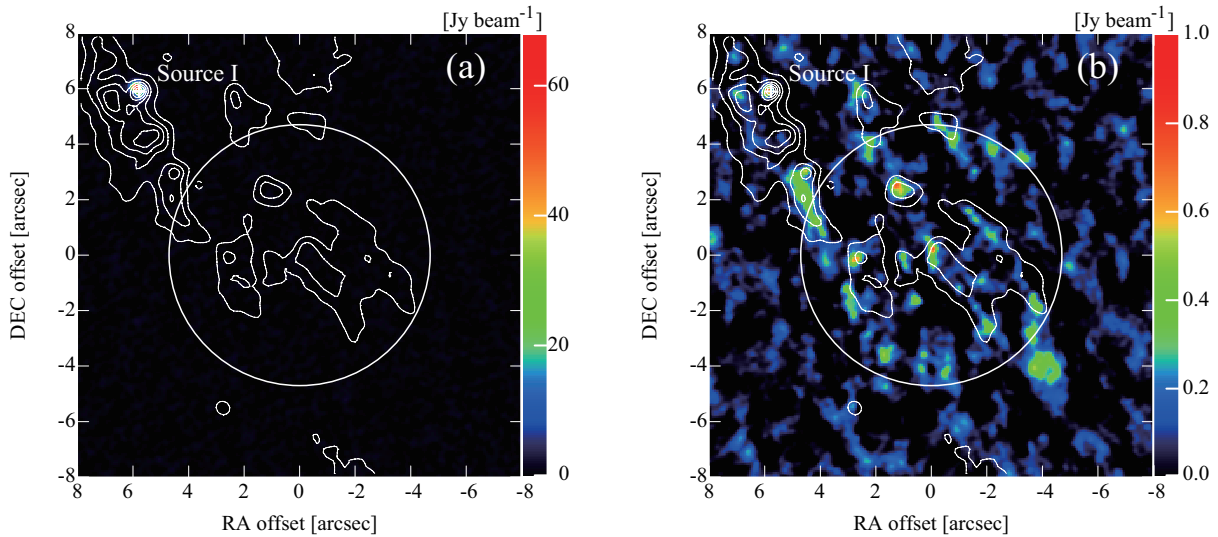}
\caption{
Maps of the Orion KL region with the position of Source I indicated. 
Contours show the distribution of ALMA band 7 (339~GHz) continuum emission \citep{hirota2015}. 
The contours start at -5$\sigma$ level with an interval of 10$\sigma$ (-5, 5, 15, 25, ...) with a 1$\sigma$ level of 9~mJy~beam$^{-1}$. 
A white circle indicates the half-power beam width (primary beam size) of an ALMA 12~m antenna at band 9 of 9.4\arcsec. 
(a) The intensity distribution of the 658~GHz H$_{2}$O line at the peak LSR (Local Standard of Rest) velocity channel of 13.0~km~s$^{-1}$ is shown by a color image. 
(b) A band 9 continuum image is shown in color. 
The primary beam attenuation for ALMA band 9 data is not corrected for in either panel. 
}
\label{fig-fov}
\end{center}
\end{figure*}

The tracking center position of Orion~KL was adopted from the 22~GHz H$_{2}$O supermaser, RA(J2000)=05h35m14s.1250, Decl(J2000)=-05d22\arcmin36\arcsec.486 \citep{hirota2011,hirota2014b}. 
Because the original target of our ALMA Cycle 0 project was the supermaser, Source~I was detected serendipitously. 
It is located at (5.8\arcsec, 5.9\arcsec) northeast of the tracking center position, and hence, outside the primary beam size of the ALMA 12~m antenna, 9.4\arcsec \ at 658~GHz (Figure \ref{fig-fov}). 
Bandpass, flux, and gain calibrators were J0423-013, Callisto, and J0607-085, respectively. 

We observed four spectral windows (spw) in band 9 with bandwidths of 938~MHz for both of the linear polarizations. 
The spw centered at 658.2~GHz included the $\nu_{2}$=1, 1$_{1, 0}$-1$_{0, 1}$ line of H$_{2}$O at a rest frequency of 658.006550~GHz \citep{chen2000}. 
We also observed other H$_{2}$O lines at 645.766010~GHz (9$_{7, 3}$-8$_{8, 0}$) and 645.950620~GHz (9$_{7, 2}$-8$_{8, 1}$) in the vibrationally ground state, with a lower state energy of $E_{l}$=2574~K \citep{chen2000}, which were in another spw centered at 646.8~GHz. 
However, we could not detect these two lines with an rms noise level (1$\sigma$) of $\sim$1~Jy~beam. 
Two other spws at 640.6~GHz and 662.2~GHz were used for continuum emission. 
The spectral resolution and number of spectral channels were 244~kHz and 3840~channels, respectively. 
The corresponding velocity resolution was 0.11~km~s$^{-1}$ at 658~GHz. 

\section{Data analysis}

Synthesis imaging and self-calibration were done using the Common Astronomy Software Applications (CASA) package in a standard manner\footnotemark \footnotetext{https://casaguides.nrao.edu/index.php/Main\_Page}. 
Part of the data plotting and image analysis were made by using the Astronomical Image Processing System (AIPS) package. 
We used the calibrated visibility data delivered by the East Asia ALMA Regional Center (EA-ARC) to make synthesized images. 
We first conducted Doppler tracking with the CASA task {\tt{cvel}}, setting the rest frequency to 658.006550~GHz. 
Next we identified line-free channels and subtracted the continuum emission from the calibrated visibility data using the CASA task {\tt{uvcontsub3}}. 

Self-calibration was done for the peak channel of the H$_{2}$O line at the LSR (Local Standard of Rest) velocity of 13.0~km~s$^{-1}$ with the CASA task {\tt{clean}} and {\tt{gaincal}}. 
First we solved only for phase solutions, with a short integration time of 5 seconds. 
We then self-calibrated the data with longer integration times (30 or 72 seconds) and solved for both phase and amplitude solutions. 
These phase and amplitude solutions were applied to rest of the spectral channels, including other spws for continuum data by using the CASA task {\tt{applycal}}. 
The resultant rms noise level for each line-free channel was 1.2~Jy~beam$^{-1}$ (1$\sigma$) at 658~GHz, although it increases to 1.6~Jy~beam$^{-1}$ (1$\sigma$) at the channels with significant emission. 
The synthesized beam size was 0.28\arcsec$\times$0.25\arcsec \ at a position angle of -64~degrees using natural weighting. 
Because the baseline length ranged from 21-385~m (45-845~k$\lambda$), extended structures of $>$5\arcsec \ are resolved out. 

Applying the self-calibration solutions to all of the spectral channels and spws, 
we produced a spectral line data cube with a channel spacing of 0.11~km~s$^{-1}$. 
For illustration purpose, we also made a lower-resolution channel map by integrating 9 velocity channels ($\sim$1~km~s$^{-1}$) as shown in Figure \ref{fig-chmap}. 
The moment 0 and 1 maps were made using the CASA task {\tt{immoments}} (Figure \ref{fig-map}). 

By using the line-free channels, we also made a synthesis image of the continuum emission of this region. 
As shown in Figure \ref{fig-fov}(b), we detected Source I and some other compact sources possibly corresponding to the main dust ridge, compact ridge, and other compact radio/infrared sources \citep{greenhill2004a, favre2011, friedel2011, sitarski2013, plambeck2013, hirota2015}. 
Interpretation is not straightforward as our image can retrieve only a small fraction of the total (large scale) continuum emission which is ``resolved out'' by ALMA. 
This also produces imaging artifacts. 
A brief summary of the continuum emission is given in Section \ref{sec-cont}, although detailed discussion is beyond the scope of this paper. 

\subsection{Flux density scale and its uncertainty
\label{sec-flux}}

Here we discuss the flux density scale and its uncertainty caused by two major factors. 
Firstly, we estimate the flux scale affected by the primary beam correction. 
The target source, Source~I is located outside the FWHM of the primary beam of the ALMA 12~m antenna of 9.4\arcsec \  as described above. 
The position offset is (5.8\arcsec, 5.9\arcsec) northeast of the tracking center position. 
If we consider a possible pointing error of 0.6\arcsec \ as described in the ALMA Cycle 0 Technical Handbook \footnotemark
\footnotetext{http://almascience.eso.org/documents-and-tools/cycle-0/alma-technical-handbook/at\_download/file}, the primary beam correction factor for Source~I is estimated to be 8.6$^{+3.3}_{-2.2}$. 
Thus, the flux uncertainty is $\sim$40\%.

Next, we consider other sources of uncertainty in the absolute flux calibration. 
We find that a flux density variation of the gain calibrator J0607-085 among four spws range from 0.64~Jy to 1.51~Jy in this epoch. 
The mean and standard deviation of the flux density of J0607-085 are 1.12~Jy and 0.36~Jy, respectively, suggesting that an additional flux calibration error of 30\%. 
This value is larger than the absolute flux calibration accuracy presented in the ALMA Cycle 0 Technical Handbook of 20\%.  
We find less flux density variation in another epoch on July 17 ranging from 0.61 to 0.66 Jy, although these data are not employed in this paper. 
Their mean and standard deviation are 0.63~Jy and 0.02~Jy, respectively. 
This value is consistent with that of the lowest value for one of the spws in Aug. 25 of 0.64~Jy. 
Because the spw including the 658~GHz H$_{2}$O line shows the highest flux density of 1.51~Jy on Aug. 25,  the flux calibration for the 658~GHz H$_{2}$O line could be overestimated by a factor of 2.4 on Aug. 25. 
The possible flux calibration error at 658~GHz could be affected by the larger atmospheric opacity close to the 658~GHz H$_{2}$O line and/or the 656~GHz O$_{3}$ line \citep{schilke2001}. 
In this paper, we employ the absolute flux density of J0607-085 to be 0.64~Jy. 
We correct this spw-to-spw flux variation for the 658~GHz line by a factor of 0.64/1.51=0.42. 
This may underestimate the flux density of the 658~GHz H$_{2}$O line by a factor of 0.42 at maximum if the actual flux density of J0607-085 is 1.51~Jy. 

In summary, the total flux calibration uncertainty is $\sim$50\% for both of the 658~GHz H$_{2}$O line and continuum data, while the absolute flux scale of the 658~GHz line could be underestimated by a factor of 0.42. 

\subsection{Absolute and relative positional accuracy}

We measured the position of Source~I using the band 9 continuum map (Figure \ref{fig-fov}) by fitting a single Gaussian with the CASA task {\tt{imfit}}. 
The measured position of Source~I is listed in Table \ref{tab-position}. 
A typical formal error in the Gaussian fitting is 5~mas as indicated in the Table \ref{tab-position}. 
We note that this does not include a systematic error term for the absolute astrometry. 
For comparison, we also list the predicted position of Source~I based on accurate absolute astrometry of the VLA 43~GHz data \citep{goddi2011}. 
The errors for the VLA data in Table \ref{tab-position} include those of proper motion and absolute positional accuracy of $\sim$5~mas \citep{goddi2011}. 
We find that the position offset of the band 9 continuum relative to the VLA 43~GHz data are 40~mas and 20~mas in right ascension and declination, respectively. 
These values could be the absolute position uncertainty in the continuum observations of Source~I at ALMA band 9. 
Possible source of astrometric error include a contribution from poorly imaged (see above) extended structure of the hot core and main dust ridge \citep{hirota2015}, uncertainties in primary beam attenuation of the images, and calibration errors. 

As discussed above, we applied the phase and amplitude calibration solutions to all spws. 
This allows us to register the 658~GHz H$_{2}$O map and that of the band 9 continuum emission. 
Thus, the error in the relative positions between the 658~GHz H$_{2}$O line and band 9 continuum should be determined by the signal-to-noise ratio and the beam size. 
As listed in Table \ref{tab-position}, the relative positional error of the ALMA band 9 continuum and the moment 0 map of the 658~GHz H$_{2}$O line is 7~mas, obtained from the formal errors in the Gaussian fitting. 

For each velocity channel map of the 658~GHz H$_{2}$O line emission, we performed a two dimensional Gaussian fit to determine peak positions using the AIPS task SAD (e.g. Figure \ref{fig-centroid}(a)). 
At the peak velocity channel of the 658~GHz H$_{2}$O line, the signal-to-noise ratio is $\sim$100, and hence, the positional precision is estimated to be $\sim$0.002\arcsec. 
Errors in the Gaussian fitting which range from 0.001\arcsec \ to 0.05\arcsec \ (1$\sigma$) with the typical value of 0.01\arcsec \ are consistent with this estimate. 
Because some of the channel maps clearly show spatial structure indicative of multiple components and/or structure larger than the beam size of $\sim$0.26\arcsec, a single Gaussian model may result in larger positional uncertainties. 

In summary, the absolute position uncertainty of Source~I in the present ALMA band 9 continuum observations is $\sim$50~mas, and the relative positional errors between the band 9 continuum and H$_{2}$O lines are 1-50~mas depending on the source structure and signal-to-noise ratio.  

\begin{deluxetable*}{llllll}
\tablewidth{0pt}
\tabletypesize{\scriptsize}
\tablecaption{Position and size of Source~I
 \label{tab-position}}
\tablehead{
\colhead{Tracer} & \colhead{$\alpha$(J2000)}  & \colhead{$\delta$(J2000)}  & 
\colhead{FWHM(\arcsec$\times$\arcsec)\tablenotemark{a}} & \colhead{PA (deg.)} & \colhead{Reference} }
\startdata
658~GHz H$_{2}$O line (moment 0) & 05h35m14s.5164(3) & -05d22\arcmin30\arcsec.491(5) & 0.525(10)$\times$0.384(9) & 44.6(12) & \\
Band 9 continuum                        & 05h35m14s.5182(4) & -05d22\arcmin30\arcsec.537(5) & 0.345(10)$\times$0.269(10) & 155(3) & \\
VLA 43~GHz continuum    & 05h35m14s.5156(4) & -05d22\arcmin30\arcsec.560(6) & \nodata & \nodata & \citet{goddi2011} 
\enddata
\tablenotetext{a}{Convolved with a synthesized beam of 0.28\arcsec$\times$0.25\arcsec \ with a position angle of -64 deg. }
\tablecomments{For ALMA band 9 data, positions and sizes are measured by a single Gaussian fitting. 
Numbers in parenthesis represent fitting errors determined by the CASA task {\tt{imfit}} in units of the last significant digit. 
Errors do not include astrometric uncertainties of 40~mas and 20~mas in right ascension and declination, respectively. 
For the VLA 43~GHz continuum, the absolute position is calculated from that in 2009 January 12 and corrected for the measured proper motion to the epoch of the ALMA data \citep{goddi2011}. 
Numbers in parenthesis represent errors that include both the absolute astrometry and proper motion uncertainties in units of the last significant digit. }
\end{deluxetable*}

\section{Results and discussion}

\subsection{Spatial structure}

\begin{figure}
\begin{center}
\includegraphics[width=8cm]{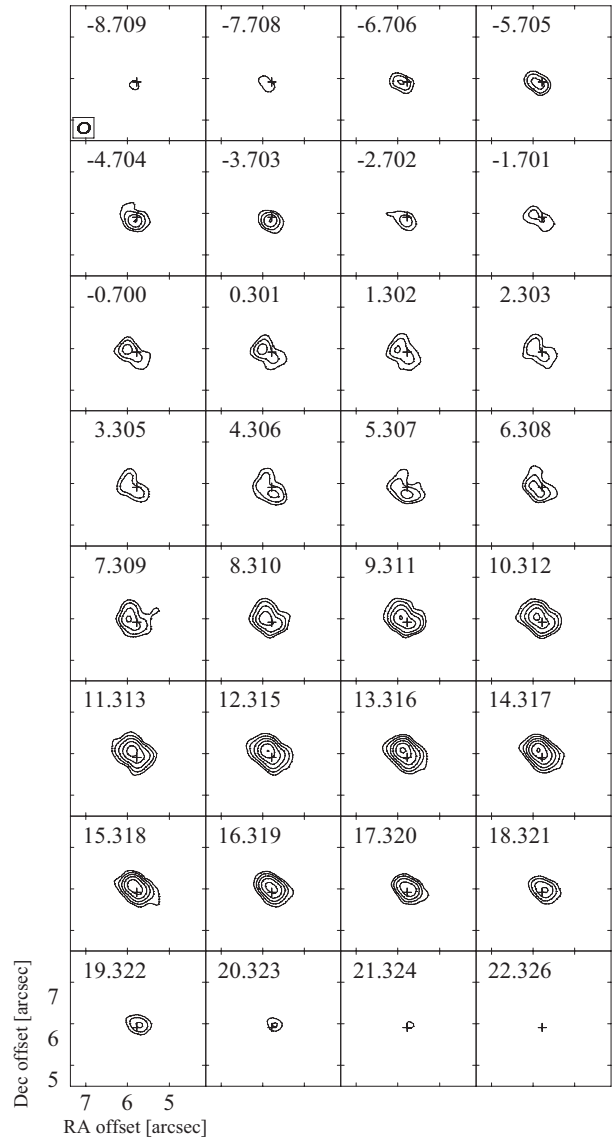}
\caption{Channel map of the 658~GHz H$_{2}$O line. 
The map is smoothed with the velocity width of 1~km~s$^{-1}$ for illustration purpose. 
The LSR velocity of the map is indicated in each panel. 
The systemic velocity of Source~I is $v_{LSR}$=5~km~s$^{-1}$. 
The contour levels are 25~Jy~beam$^{-1}$ (1$\sigma$) times -5, 5, 10, 20, 40, 80, and 160. 
A cross represents the position of the band 9 continuum emission peak of Source~I. 
A synthesized beam, 0.28\arcsec$\times$0.25\arcsec \ with a position angle of -64~degrees, is shown at the bottom-left corner of the first panel (-8.7~km~s$^{-1}$). 
Corrections for primary beam attenuation and spw-to-spw flux variation have been applied. }
\label{fig-chmap}
\end{center}
\end{figure}

Figure \ref{fig-chmap} shows channel maps of the 658~GHz H$_{2}$O line smoothed to a velocity width of 1~km~s$^{-1}$. 
The systemic velocity of Source~I is $v_{LSR}$=5~km~s$^{-1}$ \citep[e.g.][]{plambeck1990}. 
As can be seen in Figures \ref{fig-fov} and \ref{fig-chmap}, the emission of the 658~GHz line is concentrated only around Source~I, and the spatial distribution is compact with a size of less than 1\arcsec. 
Each channel map reveals a slightly more extended feature than that of the beam size, $\sim$0.26\arcsec \ or 100~AU. 
The compact distribution toward Source~I is similar to those observed for the 321~GHz and 336~GHz H$_{2}$O lines \citep{hirota2014a}. 
In contrast, it is quite different from that of the well studied 22~GHz H$_{2}$O maser and the 325~GHz submillimeter H$_{2}$O maser, which are distributed along a bipolar outflow on a much larger scale of $>$1000~AU \citep{genzel1981, gaume1998, hirota2007, niederhofer2012b, greenhill2013, neufeld2013}. 
This is because the 22~GHz and 325~GHz H$_{2}$O masers have the significantly different excitation requirements with the lower state energy levels of 642~K and 454~K, respectively, much lower than the values for the 658, 321 and 336 GHz lines \citep{chen2000}. 
The higher excitation lines at 658~GHz, 321~GHz, and 336~GHz trace a denser and hotter region close vicinity to Source~I. 

\begin{figure}
\begin{center}
\includegraphics[width=8cm]{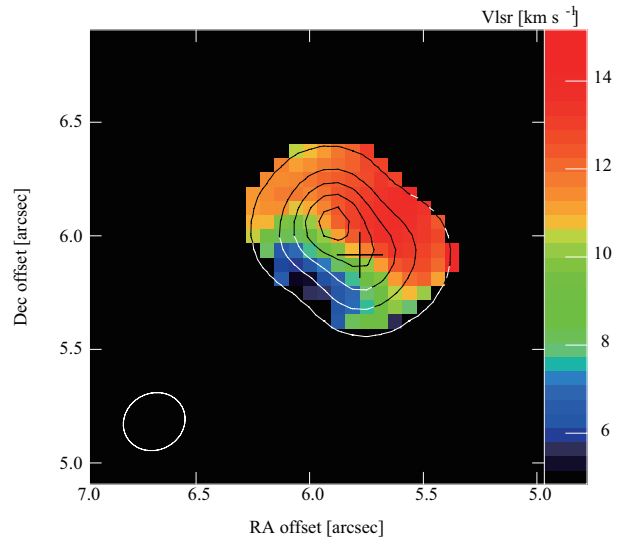}
\caption{
Moment 0 (contour) and 1 (color) maps of the 658~GHz H$_{2}$O line emission. 
The contour levels are 10\%, 30\%, 50\%, 70\%, and 90\% of the peak intensity of 3923~Jy~beam$^{-1}$~km~s$^{-1}$. 
A black cross corresponds to the position of Source~I determined by the continuum peak position. 
A synthesized beam size is shown in the bottom-left corner. 
Primary beam attenuation and spw-to-spw flux variation factors have been corrected. 
The systemic velocity of Source~I is $v_{LSR}$=5~km~s$^{-1}$, which is offset from the central velocity of the plotting range (i.e. 10~km~s$^{-1}$ as indicated by green). 
}
\label{fig-map}
\end{center}
\end{figure}

Figure \ref{fig-map} shows the moment 0 and 1 maps of the 658~GHz H$_{2}$O line. 
The spatial distribution is elongated along the northeast-southwest direction with the size of 0.525\arcsec$\times$0.384\arcsec \ or $\sim$220~AU$\times$160~AU. 
The elongation of the moment 0 map indicates that the 658~GHz H$_{2}$O line should be related to the low-velocity (18~km~$^{-1}$) outflow along the same direction. 
A velocity gradient is clearly seen perpendicular to the elongation, similar to the vibrationally excited SiO masers \citep{kim2008, matthews2010}. 
These characteristics are also analogous to that of the 321~GHz H$_{2}$O maser line \citep{hirota2014a}. 
The position angle of the moment 0 map of the 658~GHz H$_{2}$O line, 44.6$\pm$1.2~degrees (Table \ref{tab-position}), is slightly smaller than that measured both from emission locus and proper motions of 
the ground vibrational state SiO masers, 56$\pm$1~degrees \citep{greenhill2013}. 
This difference is probably due to the asymmetric brightness distribution of the 658~GHz H$_{2}$O line, where the red-shifted component of the 658~GHz H$_{2}$O line is dominant compared with the blue-shifted one (as seen in the channel map in Figure \ref{fig-chmap}). 

\begin{figure*}
\begin{center}
\includegraphics[width=16cm]{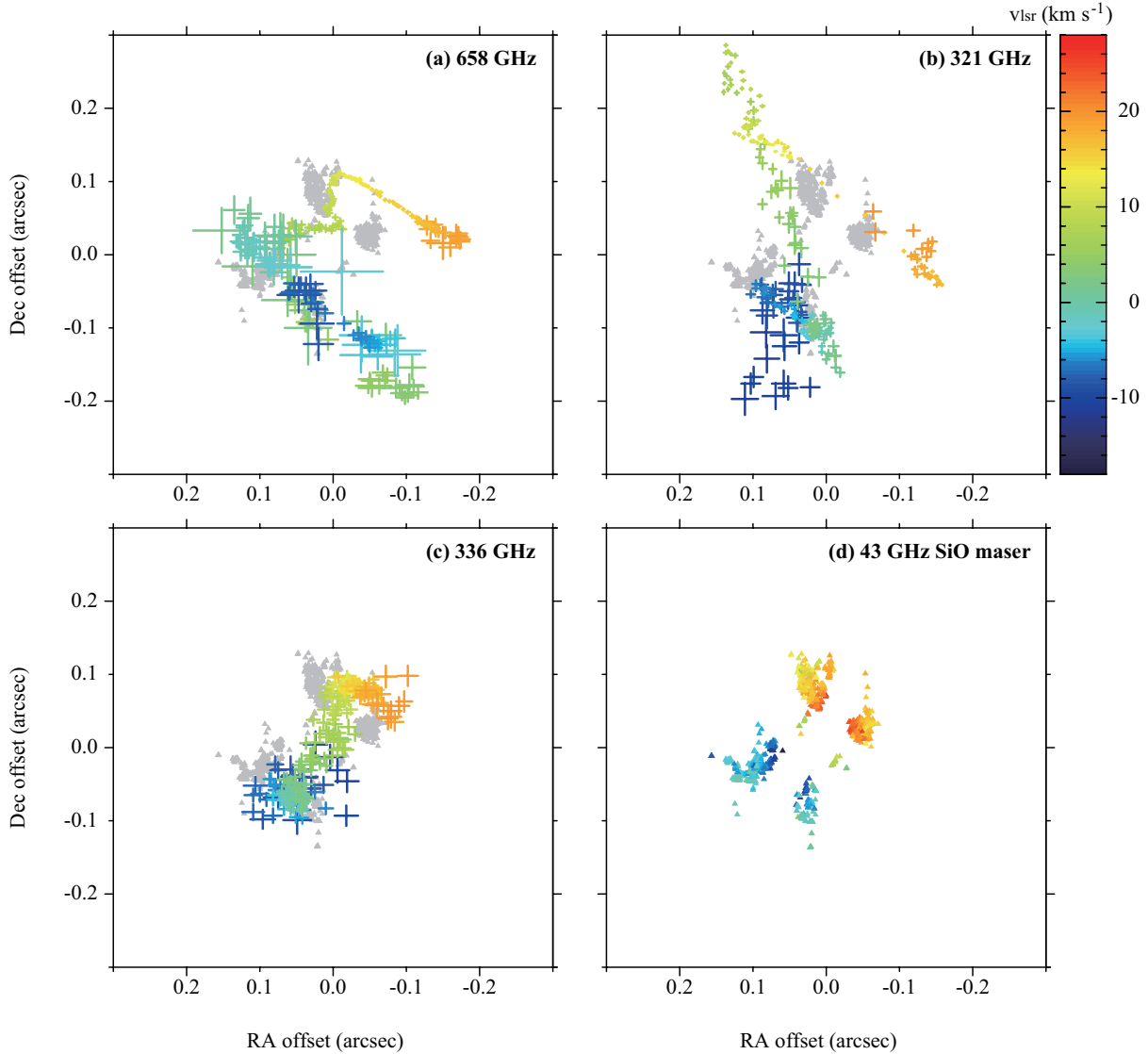}
\caption{
Velocity centroid maps of (a) the 658~GHz H$_{2}$O line, (b) the 321~GHz H$_{2}$O line, (c) the 336~GHz H$_{2}$O line, and (d) the 43~GHz SiO masers \citep{hirota2014a}. 
Colors indicate the LSR velocity. 
The systemic velocity of Source~I is $v_{LSR}$=5~km~s$^{-1}$. 
Each cross indicates a peak position with its formal error from Gaussian fitting. 
Triangles represent the positions of the SiO maser spots \citep{kim2008}. 
In panels (a), (b), and (c), the SiO maser positions are indicated by gray symbols. 
}
\label{fig-centroid}
\end{center}
\end{figure*}

We present a velocity centroid map of the 658~GHz H$_{2}$O line (Figure \ref{fig-centroid}) to compare with those of other submillimeter H$_{2}$O lines and the 43~GHz SiO masers \citep{kim2008, hirota2014a}. 
All of the map centers are registered to the peak positions of the continuum emission of Source~I at each frequency band. 
The spatial distribution of the 658~GHz H$_{2}$O line map shows an ``M-shaped'' distribution with a position angle of $\sim$45~degrees as seen in Figure \ref{fig-centroid}(a) and is slightly different from those of the 321~GHz H$_{2}$O maser line \citep{hirota2014a} and SiO masers observed with connected arrays \citep{menten1995,wright1995,baudry1998,goddi2009,niederhofer2012a}. 
However, the overall shape of the 658~GHz H$_{2}$O map is not very different from that of the 321~GHz maser line, showing two parallel structure along the nortwest-southeast directions. 
On the other hand, the velocity centroid map of the 658~GHz H$_{2}$O line is strikingly different from that of the 336~GHz vibrationally excited H$_{2}$O line \citep{hirota2014a}, which shows a linear spatial distribution perpendicular to the two ridge-like structure of the 658~GHz and 321~GHz H$_{2}$O maps. 
Between the two ridge-like structure, there is a cluster of the SiO masers detected by VLBI \citep{kim2008}. 
The sizes of the 321~GHz, 336~GHz, and 658~GHz H$_{2}$O line emission, as well as the vibrationally excited SiO masers, are comparable with each other.

The distributions of the submillimeter H$_{2}$O lines in Source~I appear to be different from those of the red supergiant VY~CMa observed with ALMA \citep{richards2014}. 
The inner rims of the 658~GHz and 321~GHz H$_{2}$O maser shells around VY~CMa are distributed at successively larger distances from the star's position, while they are outside the SiO maser shell \citep{richards2014}. 
Furthermore, the size of the outer rim of the distribution of the 321~GHz H$_{2}$O maser emission is larger than that of the 658~GHz masers by a factor of 2 (200~mas and 500~mas, respectively). 
Such different distributions are not apparent in Source~I. 
Although the angular resolution of their observations, (0.18\arcsec$\times$0.09\arcsec \ and 0.088\arcsec$\times$0.044\arcsec \ at 321/325~GHz and 658~GHzm respectively) are smaller than that of our observation, the spatial resolutions on a linear scale are comparable, $\sim$100~AU. 
We cannot rule out a possibility that the apparent structure in the velocity centroid map of Source~I could be an artifact caused by the insufficient spatial resolution as suggested by the SiO maser data \citep{goddi2009}. 
Further observational studies at higher spatial resolution with ALMA are crucial to reveal the smaller scale H$_{2}$O maser structure of Source ~I.  

\subsection{Spectral profile}

\begin{figure}
\begin{center}
\includegraphics[width=7cm]{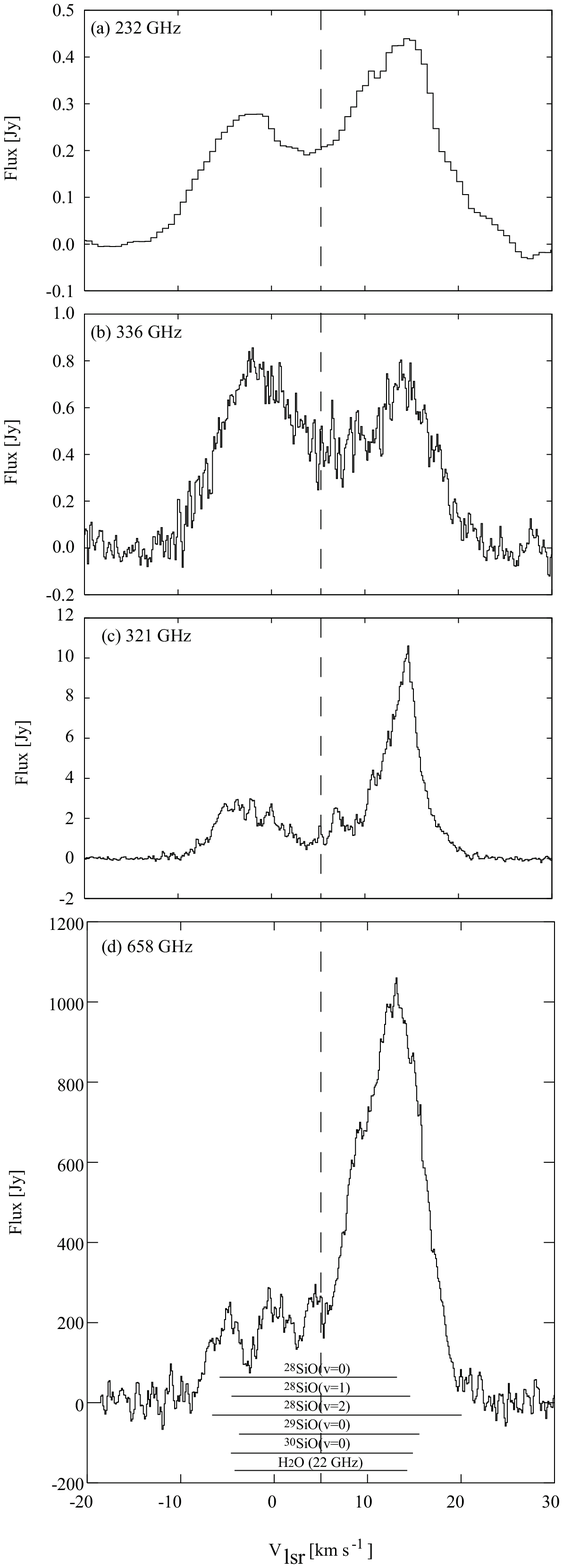}
\caption{
Spectra of the millimeter and submillimeter H$_{2}$O lines. 
(a) 232~GHz ($\nu_{2}$=1, 5$_{5,0}$-6$_{4,3}$) line taken from the ALMA SV data at band 6 \citep{hirota2012}. 
(b) 336~GHz ($\nu_{2}$=1, 5$_{2,3}$-$6_{1,6}$) line observed in ALMA Cycle 0 at band 7 \citep{hirota2014a}. 
(c) 321~GHz ($10_{2,9}$-$9_{3,6}$) line observed in ALMA Cycle 0 at band 7 \citep{hirota2014a}. 
(d) 658~GHz ($\nu_{2}$=1, 1$_{1, 0}$-1$_{0, 1}$) line. 
Primary beam attenuation and spw-to-spw flux variation factors have been corrected. 
The velocity ranges between the blue-shifted and red-shifted peaks for SiO \citep{goddi2009} and 22 GHz H$_{2}$O \citep{gaume1998} spectra are shown by horizontal bars. 
The systemic velocity of Source~I is $v_{LSR}$=5~km~s$^{-1}$, as indicated by a vertical dashed line. 
}
\label{fig-sp}
\end{center}
\end{figure}

Figure \ref{fig-sp} shows the spectral profile of the 658~GHz H$_{2}$O line along with other H$_{2}$O lines \citep{hirota2012, hirota2014a}. 
The peak flux density is 940~Jy at the LSR velocity of 13.0~km~s$^{-1}$. 
The source size of the 658~GHz H$_{2}$O channel map at 13.0~km~s$^{-1}$ is derived to be 0.42\arcsec$\times$0.31\arcsec \ by the two-dimensional Gaussian fitting of the image. 
The brightness temperature of the 658~GHz H$_{2}$O line is calculated to be 2$\times$10$^{4}$~K or higher, as the emission region is unresolved with the ALMA beam. 
Even this value may be underestimated by a factor of $1/0.42$ as discussed in Section \ref{sec-flux}. 
The observed high brightness temperature implies that maser action as a possible emission mechanism for the 658~GHz H$_{2}$O line. 

The 658~GHz H$_{2}$O line has been detected by the 10.4~m single-dish telescope of the CSO \citep{schilke2001} with the beam size of 10\arcsec. 
The main-beam brightness temperature of the CSO observation is $T_{R}^{*}$=8.9~K corresponding to the total flux denity of 300~Jy. 
This value is smller than our ALMA results by a factor of 3. 
The difference could be explained neither by the calibration error ($\sim$50\%) and pointing offset of the CSO data\footnotemark \footnotetext{The CSO observation by \citet{schilke2001} was carried out toward the hot core position at the position of RA(J2000)=05h35m14s.5, Decl(J2000)=-05d22\arcmin31\arcsec. } nor  the calibration uncertainty of the ALMA data. 
It may suggest a flux variation of the 658~GHz H$_{2}$O maser line, although further accurate flux monitoring with ALMA are still necessary to confirm this hypothesis. 

For comparison, we also present the other H$_{2}$O lines detected with previous ALMA observations \citep{hirota2012, hirota2014a} in Figure \ref{fig-sp}.  
Double-peaked spectral profiles showing brighter red-shifted component are common for all lines, except the 336~GHz line which shows a symmetric line profile. 
Nevertheless, the peak velocities of these double-peaked components are almost the same for all the lines. 
These double-peaked structures are also consistent with those of various SiO maser lines detected with the VLA \citep{goddi2009}, as shown in Figure \ref{fig-sp}. 
The observed line profiles, along with the similar distribution of the H$_{2}$O lines at 232~GHz, 321~GHz, and 658~GHz, support the idea that these lines are emitted under similar excitation conditions. 

\subsection{Excitation condition of the 658~GHz H$_{2}$O line}

Based on the spatial structure and spectral profiles, it is most likely that the 658~GHz H$_{2}$O line is associated with the base of the outflow as revealed by VLBI observations of SiO masers \citep{greenhill2004b, kim2008, matthews2010} and the 321~GHz H$_{2}$O maser line \citep{hirota2014a}. 
The brightness temperature of the 658~GHz H$_{2}$O line, $>$2$\times$10$^{4}$~K, would indicate that the line is emitted via maser action. 
In contrast, the 336~GHz H$_{2}$O line shows a different distribution (Figure \ref{fig-centroid}(c)) and a spectral profile with symmetric double-peaks (Figure \ref{fig-sp}). 
As discussed in \citet{alcolea1993} and \citet{hirota2014a}, the 336~GHz H$_{2}$O line could be a thermal line and trace the edge-on disk, unlike emission from the other transitions at 232~GHz, 321~GHz, and 658~GHz. 
We note that \citet{menten2006} the 336~GHz H$_{2}$O line toward the red supergiant VY~CMa also favor a thermal nature for this line's excitation. 

Recently, theoretical model calculations for the excitation of the 658~GHz H$_{2}$O line have been presented \citep{richards2014, nesterenok2015} for the red supergiant VY~CMa and for asymptotic giant branch (AGB) stars. 
The 658~GHz H$_{2}$O maser is predicted to be pumped efficiently for an H$_{2}$ densities of 10$^{9}$-10$^{10}$~cm$^{-3}$, a fractional H$_{2}$O abundance of 10$^{-4}$-10$^{-5}$, and gas temperatures of 1000-1500~K \citep{nesterenok2015}. 
Although the physical and chemical properties of the 658~GHz H$_{2}$O maser emitting region associated with Source~I would be expected to be different from that of a late-type star, these parameters are roughly consistent with those for the disk-outflow system of Source~I associated with SiO masers and other submillimeter H$_{2}$O lines \citep{reid2007, goddi2009, goddi2011, hirota2014a, hirota2015}. 

\subsection{Relationship between the 22~GHz H$_{2}$O supermaser}

We note that our ALMA observation was done during an active phase of the 22~GHz H$_{2}$O maser burst in Orion~KL, called a ``supermaser'' \citep{hirota2011, hirota2014b}. 
The 22~GHz supermaser is located at the phase tracking center of the ALMA observation, corresponding to the Compact Ridge region \citep{hirota2011, hirota2014b}. 
While the total flux density of the 22~GHz H$_{2}$O supermaser was decreasing on August 25, 2012, it was still $\sim 2\times10^{4}$~Jy \citep{hirota2014b}. 
In spite of the extremely high flux density of this maser, there was no significant emission of the 658~GHz H$_{2}$O line in the Compact Ridge region in our ALMA data with the 1$\sigma$ noise level of 2~Jy~beam$^{-1}$ in the relevant region. 
Thus, the 658~GHz H$_{2}$O line is not related to the 22~GHz supermaser phenomenon, as has also been reported for other transition at 321~GHz and 336~GHz \citep{hirota2014b}. 

\subsection{Continuum emission at ALMA band 9}
\label{sec-cont}

To measure the position of the 658~GHz H$_{2}$O emission, we also made a map of the ALMA band 9 continuum by applying the self-calibration solutions.  
Continuum emission is detected at the position of Source~I, although it is outside the primary beam of the 12~m antenna. 
The flux density and the peak intensity are derived from Gaussian fitting, 1.04$\pm$0.04~Jy and 0.77$\pm$0.03~Jy~beam$^{-1}$, respectively (Figure \ref{fig-fov}(b)). 
If we correct the primary beam attenuation by a factor of 8.6, the absolute flux density is calculated to be 8.9~Jy, with an uncertainty of 50\%. 
This value is slightly larger than that of the SMA observation at 690~GHz, 6.7$\pm$3.2~Jy \citep{beuther2006}, but within the mutual errors. 
The apparent source size is almost comparable to the beam size as listed in Table \ref{tab-position}, and hence, the continuum emission of Source~I is unresolved with our ALMA observation. 
Due to the large ($\sim$50\%) flux uncertainty of Source~I (which is outside the primary beam), the limited field of view, and insufficient image quality in particular for spatially extended components, we will not discuss detailed properties of the continuum emission of Source~I. 

\section{Summary}

By using ALMA Cycle 0 data, we detect the 658~GHz H$_{2}$O line toward the massive protostar candidate Source~I in the Orion KL region. 
The 658~GHz H$_{2}$O line is emitted from a compact structure with the size of $\sim$100~AU. 
The source structure is elongated along the northeast-southwest direction parallel to the low-velocity (18~km~s$^{-1}$) molecular outflow. 
A velocity gradient can be seen along the northwest-southeast direction, which is perpendicular to the outflow axis and the source elongation. 
The spectral profile represents a double-peaked structure, similar to the other millimeter/submillimeter H$_{2}$O lines at 232~GHz and 321~GHz lines \citep{hirota2012, hirota2014a}.  
The high brightness temperature of $>$2$\times$10$^{4}$~K is consistent with maser emission. 
These basic properties would suggest that the 658~GHz H$_{2}$O line is emitted from the base of the northeast-southwest low-velocity (18~km~s$^{-1}$) molecular outflow. 
Further multi-transition studies of the H$_{2}$O lines with ALMA at higher spatial resolution should provide information about the excitation mechanism and physical properties of the millimeter and submillimeter H$_{2}$O line emissions. 
They will be powerful tools to investigate dynamics, physics and chemistry in outflow launching regions associated with accretion disks around massive YSOs, which is a key issue to understand formation mechanisms of massive YSOs. 

\bigskip
We acknowledge Karl M. Menten and Mark J. Reid for a critical reading of the manuscript. 
This paper makes use of the following ALMA data: ADS/JAO.ALMA\#2011.0.00199.S. 
ALMA is a partnership of ESO (representing its member states), NSF (USA) and NINS (Japan), together with NRC (Canada) and NSC and ASIAA (Taiwan), in cooperation with the Republic of Chile. 
The Joint ALMA Observatory is operated by ESO, AUI/NRAO and NAOJ. 
T.H. is supported by the MEXT/JSPS KAKENHI Grant Numbers 21224002, 24684011, and 25108005, and the ALMA Japan Research Grant of NAOJ Chile Observatory, NAOJ-ALMA-0006. 
M.H. is supported by the MEXT/JSPS KAKENHI Grant Numbers 24540242 and 25120007.

\end{document}